# A Noise Addition Scheme in Decision Tree for Privacy Preserving Data Mining

Mohammad Ali Kadampur, Somajajulu D.V.L.N.

**Abstract**—Data mining deals with automatic extraction of previously unknown patterns from large amounts of data. Organizations all over the world handle large amounts of data and are dependent on mining gigantic data sets for expansion of their enterprises. These data sets typically contain sensitive individual information, which consequently get exposed to the other parties. Though we cannot deny the benefits of knowledge discovery that comes through data mining, we should also ensure that data privacy is maintained in the event of data mining. Privacy preserving data mining is a specialized activity in which the data privacy is ensured during data mining. Data privacy is as important as the extracted knowledge and efforts that guarantee data privacy during data mining are encouraged. In this paper we propose a strategy that protects the data privacy during decision tree analysis of data mining process. We propose to add specific noise to the numeric attributes after exploring the decision tree of the original data. The obfuscated data then is presented to the second party for decision tree analysis. The decision tree obtained on the original data and the obfuscated data are similar but by using our method the data proper is not revealed to the second party during the mining process and hence the privacy will be preserved.

——————————— ◆ ———————————

## 1 INTRODUCTION

Organizations in the modern business world collect lots of data and they are data rich. On top of data they place a layer of data mining algorithms which help extract patterns/classes/associations in the data without any *apriori* hypothesis. This process of knowledge discovery has multifold benefits to the organizations and organizations continue to plunder the data by using various mining techniques. It is imperative that once the organizations start sharing the data during the mining process the data proper gets exposed and privacy of individual records is breached. It is important there fore to mine the data without revealing the data proper to other parties. Privacy preserving data mining is such a specialized set of data mining activity where techniques are evolved to protect the privacy of the data and at the same time the knowledge discovery process is carried out without ban or bash. It is a matter of history now that Privacy Preserving Data Mining (PPDM) techniques have bailed out the Data Mining (DM) technology from a total ban .PPDM is an active field of research in knowledge engineering. In this paper we propose a novel method that ensures the data privacy in the event of decision tree analysis on the data. It is basically a noise addition framework specifically tailored toward classification task in data mining. The method also preserves averages and few other statistical parameters thus making the modified data set useful for both statistical and data mining purposes .Let D be the original data set and T be its decision tree. Our effort will be to modify D to D' and get the corresponding decision tree T' such that T and T' are similar. This way we can suggest to expose D' instead of D for mining purposes where by one can avoid revealing the data proper and still be assured of unaffected data mining results (*specifically decision tree*).

## 2 PREVIOUS WORK

For the past few years, several approaches have been proposed in the context of privacy preserving data mining. These techniques can be classified based on the different protection methods used, such as Data modification methods, Cryptographic methods with distributed privacy and Query auditing. Fig-1 shows the classification.

Data modification techniques modify the data before releasing it to the users. Data is modified in such a way that the privacy is preserved in the released data set. Several data modification techniques are proposed including noise addition[ 1][5][23] , data swapping[24 ][25], aggregation[26 ][27 ], suppression [28][29] and signal transformation[30 ][ 31].

Cryptographic methods [22] encrypt the data with encryption schemes while still allowing the data mining tasks. These methods use certain set of protocols such as secured multiparty computation(SMC).SMC techniques are not supposed to disclose any new information other than the final result of the computation to a participating party. SMC techniques are applied to distributed data sets.

---

- *Mohammad Ali Kadampur is with the Department of Computer Science and Engineering, National Institute of Technology Warangal A.P. India-506004..*
- *Somayajulu D.V.L.N is with the Department of Computer Science and Engineering, National Institute of Technology Warangal. A.P. India 506004.*

As an extension, SMC protocol is applied to centralized data sets by partitioning the data sets either vertically or horizontally.[ 32][ 33][34]. Cryptographic methods bring in the overhead of encryption decryption and are less efficient for larger data set and where data utility is of concern. Query auditing methods preserve privacy by modifying or restricting the results of a query. [35][36][37]. In these methods too many denials to a query leads to less utility of the data set. Lesser denial though increases the utility but sacrifices privacy.

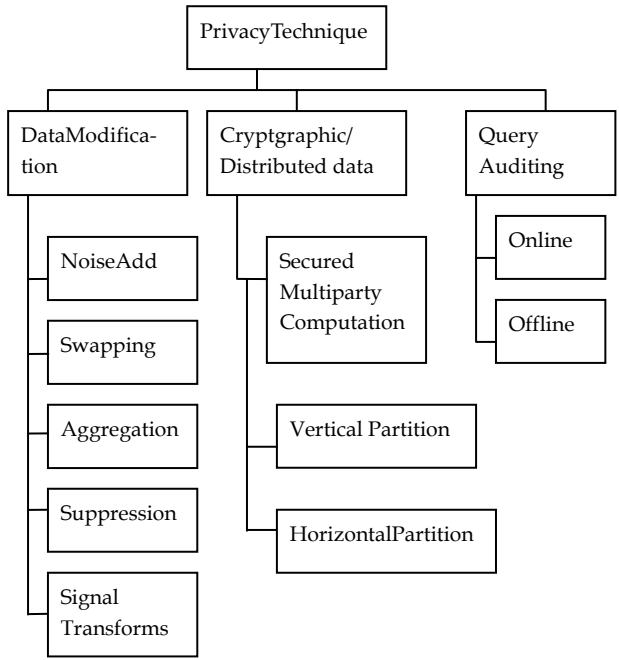

Fig-1: Classification of Privacy Preserving Data Mining Techniques.

Noise addition methods add some random number (noise) to numerical attributes. This random number is generally drawn from a normal distribution with zero mean and a small standard deviation. Noise addition to categorical values is not straightforward.

Data swapping interchanges attribute values among different records. Similar attribute values are interchanged with higher probability. All original values are kept within the data set and only the positions are swapped.

Aggregation refers to grouping. Here in these methods few records are grouped and replaced by a group representative such as in case of *income* attribute, instead of individual *income* values they can be grouped into, *high low* and *medium income*. Suppression refers to replacing an attribute value in one or more records by a missing value.

Signal Transform methods use Fourier Transformation and Wavelet transformation to modify the data. These methods are fast with improved time complexity than their predecessors. Literature of privacy preserving data mining is rich in explaining each of these methods.

However no method is complete and satisfactory. Each method suffers from one or the other kind of bias [4].If we consider the data mining tasks and classify the privacy preserving methods then another set of interesting classification can be seen such as (i) Methods that preserve statistical parameters. (ii) Methods that preserve classification results. (iii) Methods that preserve clustering results. (iv) Methods that preserve association rules. (iv) Methods that preserve more than one data mining features.

It is found that each method that is developed by research, preserves privacy for either classification results [5] [6] or clustering results [7] [8] [9] [10] or association rules [11] [12] or for combination of few data mining tasks. As the mining algorithms are improving the newer biases such as Data Mining biases (DM bias) are being defined. The challenges that the data miner faces are ever increasing with the size and complexity of data [12].

## 3 OVERVIEW OF DECISION TREE

Decision Tree algorithm uses a splitting criterion based on the information gain in the attribute. The attribute with highest information gain will form the root of the tree and algorithm iteratively continues splitting the data to form a decision tree .It essentially finds the best splitting attribute and the best splitting point of the numeric continuous attributes. It was first proposed by Quinlan in [17] and later improved to be known as C4.5 and C5.0 algorithm. The formulas of information gain and gain ratio are as follows.

$S$ is the training set, $|S|$ is the number of instances in $S$ . $|S_i|$ is the number of instance in the category $S_i$ . $freq(C_i, S)$ is the number of instances that belong to class $i$ where $i$ varies from 1 to n and $Test_A$ is the test attribute.

$$Info(S) = -\sum_{i=1}^{n}\left(\frac{freq(C_i, S)}{|S|}\right) \times \log_2\left(\frac{freq(C_i, S)}{|S|}\right)$$
(1)

$$Info_{Test_A}(S) = \sum_{i=1}^{n}\frac{|S_i|}{|S|} \times Info(S_i)$$
(2)

$$gain(Test_A) = Info(S) - Info_{Test_A}(S)$$
(3)

$$splitInfo(Test_A) = -\sum_{i=1}^{n}\frac{|S_i|}{|S|} \times \log_2\left(\frac{|S_i|}{|S|}\right)$$
(4)

$$gainRatio(Test_A) = \frac{gain(Test_A)}{splitInfo(Test_A)}$$
(5)

With respect to the following table-1 we show how the Information gain terms are calculated in the decision tree formation.Fig-1 shows the typical decision tree obtained for the sample table taken in table-2.

TABLE-1
SAMPLE DATA SET(Liver dataset)

| Liver Size | Patient's Weight | Eats Pizza | Diagnostic Class |
|---|---|---|---|
| NORMAL | 70 | YES | CLASS1 |
| NORMAL | 90 | YES | CLASS2 |
| NORMAL | 85 | NO | CLASS2 |
| NORMAL | 95 | NO | CLASS2 |
| NORMAL | 70 | NO | CLASS1 |
| ENLARGED | 90 | YES | CLASS1 |
| ENLARGED | 78 | NO | CLASS1 |
| ENLARGED | 65 | YES | CLASS1 |
| ENLARGED | 75 | NO | CLASS1 |
| SHRINKED | 80 | YES | CLASS2 |
| SHRINKED | 70 | YES | CLASS2 |
| SHRINKED | 80 | NO | CLASS1 |
| SHRINKED | 80 | NO | CLASS1 |
| SHRINKED | 96 | NO | CLASS1 |

In the dataset in table-1 nine samples belong to CLASS1 and five samples belong to CLASS2. Let $Test_A$ ,is the *LiverSize* attribute, the entropy before splitting is

$$Info_{Test_A}(S) = \frac{5}{14}\left(-\frac{2}{5}\log_2\left(\frac{2}{5}\right) - \frac{3}{5}\log_2\left(\frac{3}{5}\right)\right)$$

$$+ \frac{4}{14}\left(-\frac{4}{4}\log_2\left(\frac{4}{4}\right) - \frac{0}{4}\log_2\left(\frac{0}{4}\right)\right)$$

$$+ \frac{5}{14}\left(-\frac{3}{5}\log_2\left(\frac{3}{5}\right) - \frac{2}{5}\log_2\left(\frac{2}{5}\right)\right)$$

$$= 0.940 \, bits$$

$$Info_{Test_A}(S) = \frac{5}{14}\left(-\frac{2}{5}\log_2\left(\frac{2}{5}\right) - \frac{3}{5}\log_2\left(\frac{3}{5}\right)\right)$$

$$+ \frac{4}{14}\left(-\frac{4}{4}\log_2\left(\frac{4}{4}\right) - \frac{0}{4}\log_2\left(\frac{0}{4}\right)\right)$$

$$+ \frac{5}{14}\left(-\frac{3}{5}\log_2\left(\frac{3}{5}\right) - \frac{2}{5}\log_2\left(\frac{2}{5}\right)\right)$$

$$= 0.694 \ bits$$

$$Gain(Test_A) = 0.940 - 0.694 = 0.246$$

If the splitting is based on the attribute *EatsPizza* a similar computation will give a Gain of 0.048 bits and for the *pa*

*tients weight* attribute the gain will be 0.103 bits. Since Gain (*LiverSize*) > Gain (*EatsPizza*) > Gain (*Patient'slWeight*) the decision tree algorithm will select LiverSize as splitting criteria. The decision tree after the first split looks as shown in Fig-2.

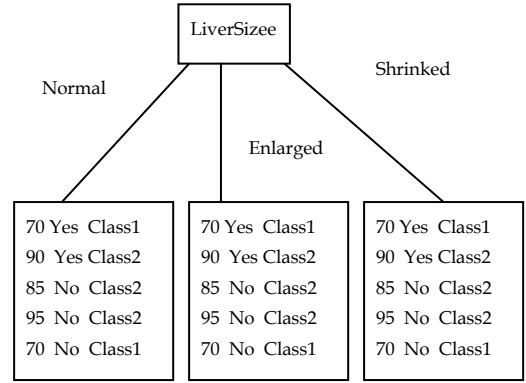

Fig-2: The Decision Tree of the Sample Data Set (Liver-DataSet) after the first split.

The algorithm is applied recursively to each child node until all samples at a node are of one class. Every path to the leaf in decision tree represents a classification rule. Attribute selection is based on minimizing an information entropy measure applied to the example at a node.

## 4 OUR APPROACH

We use Quinlan's [17] C5.0 decision tree builder on the selected data set [18] and obtain the decision tree of the original data set. We then approach a unique method of listing the nodes (attributes) that we touch in the path from the root of the tree to the leaf. We then use a noise addition strategy for each of the attributes.

### 4.1 Terminology

In any decision tree we have some leaves and some internal nodes. The path that leads from the root to the leaf is called Leaf Reaching Path(LRP) and nodes that form LRPs are listed as Leaf Reaching Path Attributes (LRPAs).For example in the tree in Fig-3, for LEAF₄ the LRPA is *percentage low income earners, Av rooms per dwelling, pupil-teacher ratio*. The path that doesn't lead to a leaf is called Leaf Wrong Path (LWP).There may be many Wrong Paths to a leaf. set of attributes that don't form LRP are grouped as Leaf Wrong Path Attributes (LWPA).The LWPA for LEAF₄ is *nitric oxides ppm* .Each leaf in the tree has a set of LRPAs and LWPAs. Each LRPA attribute or LWPA attribute may be numerical or categorical. If the attribute is categorical (either in LRPA list or LWPA list), we are using a CAPT, (Categorical Attribute Perturbation Technique) for perturbing it. For numerical attributes of LRPA & LWPA we will use specific noise addition techniques (PTLRPA & PTLWPA) explained in the sections that follow.

PTLRPA is the perturbation technique used to perturb numeric attributes of LRPA and PTLWPA is the perturbation technique used for perturbing numeric attributes of LWPA.

We use a wrapping function V_WRAP to wrap the numeric values if, after addition of noise values exceed their respective attribute domain. Domain of the attribute is the range of values for that attribute. Example, for an attribute such as *age* the domain would be [1..100].

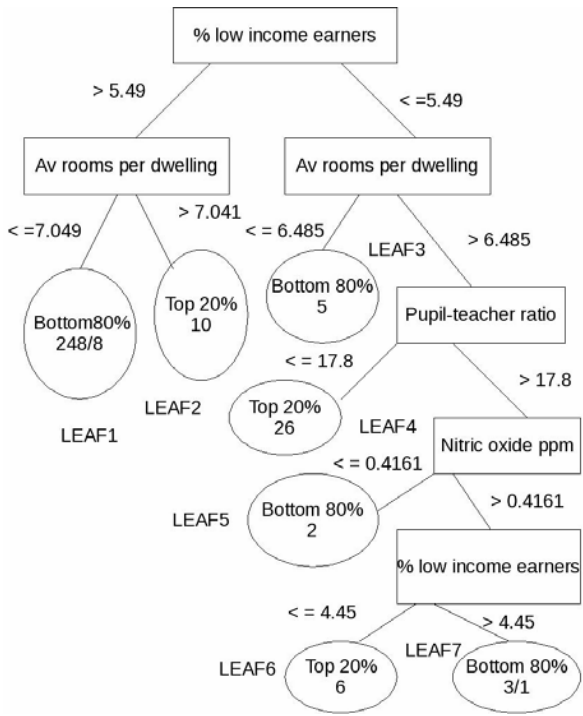

Fig-3 Decision Tree from BHP Dataset used for explaining terminology of our approach.

## 4.2 CAPT

Categorical Attribute Perturbation Technique basically shuffles/changes the attribute values with certain probability. Depending upon the type of the leaf in the tree. The leaf can be a heterogeneous leaf or a homogeneous leaf. Heterogeneous leaf is the one that contains more than one class type and homogeneous leaf is the one that contains only one class type Heterogeneous leaf will have some majority records and zero or more minority records. majority records are those whose occurrence is maximum times than the other records with respect to the class identified. In *Fig-4*, $L_1$ is a heterogeneous leaf and $L_2$ is a homogeneous leaf. In $L_1$ A is the majority class and S is the minority class.

Let *p* be the user defined probability, then CAPT shuffles the class values of the leaf with probability *(1-p)*.

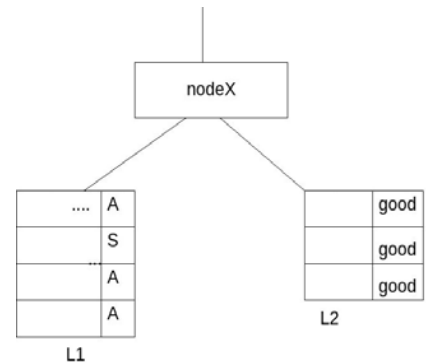

Fig.4 : Heterogeneous and homogeneous leaves.

## 4.3 CAPT Algorithm

BEGIN CAPT
    Scan the records one by one.
    For each record
    DO
    Identify the leaf to which
    the record belongs.
    IF (the leaf has no siblings)
    DO
        IF (leaf is heterogeneous)
        Assign,
        m=The number of majority records.
        $n_i$=the number of records having
        minority i[th] class.
        t=Number of different minority classes.
        $k = \sum_{i=1}^{t} n_i$ ;
        q=m/m+k ;
        $l_i$=$n_i$/$n_i$+k ;
        p=user defined probability.
            BEGIN shuffling the records
            FOR(the majority class records)
            assign (1-p)*q probability ,
            FOR(minority class records)
            assign (1-p)*$l_i$  probability.
            END shuffling the records
        END IF
    END DO
    ELSE
        IF(Leaf has siblings)
        Identify the majority class in the leaf.
        Assign,
        New_class_of the record
        = majority_class_of_the_leaf.
        END IF
    END DO
END CAPT.

## 4.4 PTLRPA Algorithm

Perturbation Technique for Leaf Reaching Path Attributes. In this technique first We obtain the attributes that are tested in reaching the leaf. If the attribute is categorical, we use CAPT otherwise we use the following PTLRPA algorithm to add noise to the numerical attributes of LRP.

```
BEGIN PTLRPA
    Find normal distributions of all numeric
    attributes of Decision Tree.
    For each record of the dataset,
    Do
            Determine the leaf L to which it belongs.
            Identify the LRPAs.
            Identify Domains of each numeric attribute
            of LRPA.
            Add a small noise drawn from respective
            distributions of attributes, having certain
            mean and variance.
            IF ( attribute value+Noise) > Domain Value
                Call a wrapper function (V_WRAP) to
                wrap around the value.
            ENDIF
    END DO
END PTLRPA.
```

## 4.5 PTLWPA Algorithm

Perturbation Technique for Leaf Wrong Path Attributes. In this technique first we obtain the attributes that are *not* tested in reaching the leaf. If the attribute is categorical, we use CAPT, otherwise we use the following PTLWPA algorithm to add noise to the numerical attributes of LWP.

```
    BEGIN PTLWPA
    Find the distributions of all numeric  attributes of
    Decision Tree.
    For each record of the dataset,
    Do
            Determine the leaf L to which it belongs.
            Identify the LWPAs.
            Identify Domains of each attribute of
            LWPA.
            Add a small noise drawn from
            respective distributions of attributes,
            having certain mean and variance.
            IF
            after addition, the attribute value
            exceeds its domain value,
            THEN
            call a wrapper function(V_WRAP)to
            wrap around the value.
            ENDIF
    END DO
    END PTLWPA
```

## 4.6 V_WRAP Algorithm

This function is called by PTLRPA or  PTLWPA algorithms when after addition of noise the attribute value exceeds its domain value.

```
BEGIN V_WRAP
    DO
        Record the Domain limits [a, (a+D)] of the attribute.
        Get the input value $P_i$ for Wrapping.
        Compare $P_i$ with max domain limit.
            IF ($P_i$ > (a+D))
                d=$P_i$-(a+D)
                ELSEIF ($P_i$ < a )
                    d=$P_i$-a
            ENDIF
                    $P_f$=a+d-1
                    RETURN($P_f$)
        END IF
    END DO
END V_WRAP .
```

## 5  DATA SETS

In our experiments we have used the following data sets from UCI machine learning repository [38].

### 5.1 The BHP data set

The *Boston Housing Price* (BHP) data set has altogether 12 attributes, out of which one is the categorical attribute with domain size 2 top 20%, bottom 80%,. The non-class attributes are *crime rate, proportion large lots, proportion industrial, nitric oxide ppm, Av rooms per dwelling, proportion pre-1940, distance to employment centers, accessibility to radial highways, property tax rate 10,000 dollars, pupil-teacher ratio and percentage low income earners}*. All non-class attributes are continuous. Other two non-class attributes "CHAS" and "B" are ignored throughout our experiments.

### 5.2 Census Income Data Set

This data set has fourteen attributes, six continuous and eight nominal. It altogether has 48842 instances and 16281 testing data instances. This data set can be downloadable from University of California Irvine repository.

### 5.3 Car Evaluation Data Set

This data set is derived from a hierarchical decision model, and was first used in[21]. This data set has mainly six categorical attributes and 1728 instances. Associated task is classification.

## 6 EXPERIMENTS AND RESULTS

We initially conducted our experiments on the sample dataset of table-1 and later the other data sets were tested We added the random noise drawn from the distributions to the *Patient'sWeight* attribute and the perturbed table was obtained (table-2). The random noise was chosen from the distribution shown in Fig-5. The decision tree obtained for these two tables, table-1 and table-2 were same.(Fig-6).

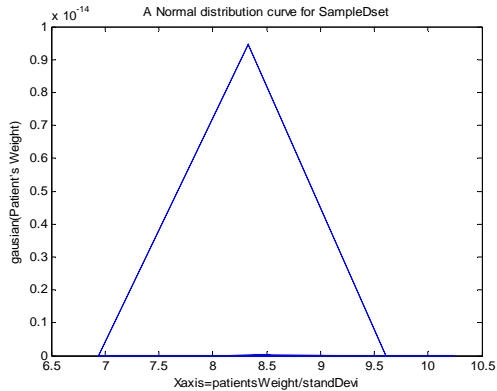

Fig-5. Normal distribution of sample data set (LiverDataSet) of table-1.

We then selected the datasets from UCI machine learning repository. We built a classifier from the perturbed tree and applied the classifier on the training and testing data sets separately. We also built a classifier from the original tree and applied on the training and testing data sets. We then compared their accuracies. (Tabl-3 ).

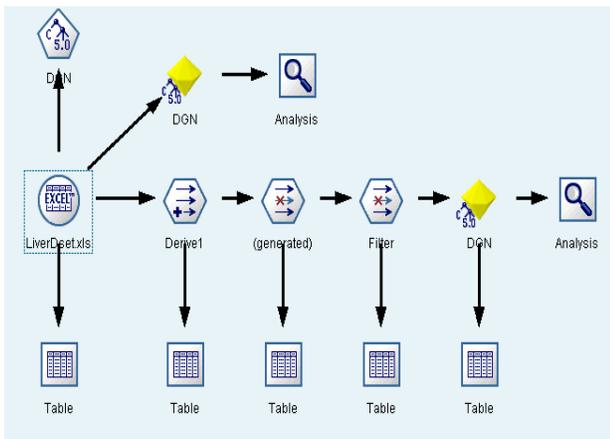

Fig-6 Experimental setup with C5.0 node.

DGN-Diagnosis.
Two instances of DGN node are shown in Fig-6.
The DGN node after Derive1 and generated nodes is receiving perturbed data as input.

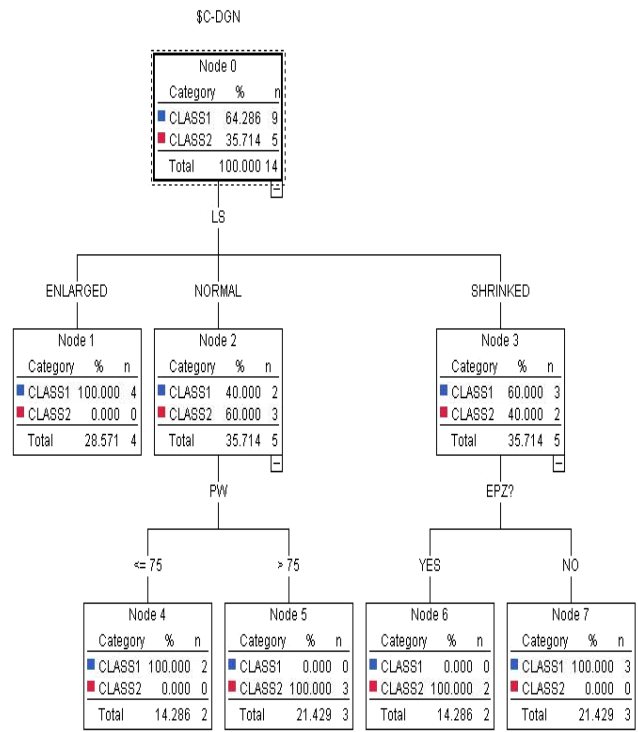

Fig-7 The Decision tree for the sample dataset (LiverDataSet) and its perturbed instance

TABLE-2
INSTANCE OF ORIGINAL TABLE WITH PERTURBED VALUES FOR THE ATTRIBUTE "Patients Weight".

| Liver Size | Patient's Weight (original) | Patient's Weight (pertbd) | Eats Pizza | Diagnostic Class |
|---|---|---|---|---|
| NORMAL | 70 | 65.74 | YES | CLASS1 |
| NORMAL | 90 | 85.74 | YES | CLASS2 |
| NORMAL | 85 | 80.74 | NO | CLASS2 |
| NORMAL | 95 | 90.74 | NO | CLASS2 |
| NORMAL | 70 | 65.74 | NO | CLASS1 |
| ENLARGED | 90 | 85.74 | YES | CLASS1 |
| ENLARGED | 78 | 73.74 | NO | CLASS1 |
| ENLARGED | 65 | 60.74 | YES | CLASS1 |
| ENLARGED | 75 | 70.74 | NO | CLASS1 |
| SHRINKED | 80 | 75.74 | YES | CLASS2 |
| SHRINKED | 70 | 65.74 | YES | CLASS2 |
| SHRINKED | 80 | 75.74 | NO | CLASS1 |
| SHRINKED | 80 | 75.74 | NO | CLASS1 |
| SHRINKED | 96 | 91.74 | NO | CLASS1 |

TABLE-3
PREDICTION ACCURACIES OF CLASSIFIER BEFORE AND AFTER PERTURBATION OF THE DATA SETS

|  | Before Perturbation | After Perturbation |
|---|---|---|
| BHP Dataset | 85.61% | 84.88% |
| CI Dataset | 84.23% | 83.85% |
| CarEvaluation Data Set | 86.67% | 86.58% |

Ideally, the accuracy of a perturbed classifier should be as good as the accuracy of the original classifier. We observe that the comparative results are tending to fulfill this requirement. The data quality of the perturbed data set is considered to be high when the perturbed data set is similar to the original data set in terms of decision tree and the classifier accuracies.

# 7 CONCLUSION

The approach taken in this paper integrates both categorical and numeric data types and focuses on privacy preserving during classification, particularly in decision tree analysis. The noise addition methods used are effective in preserving the privacy of the data proper and producing prediction accuracies on par with the original dataset. Crucial properties of a noise addition technique are the ability to maintain data quality and ensure individual privacy. More experiments are to be conducted on data quality and security level measurements. In the context of various data mining tasks, as our approach deals only with the classification task, we conclude that our approach is addressing the issue of PPDM partially. Future research however may incorporate the approach taken in this paper to evolve a unified privacy preserving framework that addresses as many data mining tasks as possible.

# REFERENCES

[1] Agrawal R.,Srikant R., ``Privacy Preserving Data Mining.,'' *In the Proceedings of the ACM SIGMOD Conference*. 2000.

[2] L.Sweenay., ``Achieving k-anonimity privacy protection using generalization and suppression.,'' *International Journal on Uncertainty, Fuzziness and Knowledge based Systems.*,10(5),pp.571-588,2002.

[3] Kun Liu., Hillol Kargupta. ``Random Projection-Based Multiplicative Data Perturbation for Privacy Preserving Distributed Data mining.,''

[4] *IEEE Transactions on Knowledge and Data Engineering.*, Vol.18,No.1,2006.

[5] K.Muralidhar.,R.Sarathy.,``A General additive data perturbation method for data base security.,'' *Journal of Management Science. A*5(10):1399-1415,2002.

[6] Li Liu., Murat Kantarcioglu.,Bhavani Thuraisingham. ``Privacy Preserving Decision Tree mining from Perturbed Data.,'' *In proceedings of 42nd Hawaii International Conference on System Sciences*. ,2009.

[7] Chen K.,Liu L. ``A random rotation perturbation approach to privacy data classification.,'' *Proceedings of IEEE Conference on Data Mining (ICDM).* pp.589-592,2005

[8] S.Merugu.,J.Ghosh., ``Privacy Preserving Distributed Clustering using generative models.,'' *In Proceedings of 3rd IEEE International Conference on Data Mining (ICDM03).* Vol.19,Issue 22,pp.211-218, 2003.

[9] S.R.M.Olivera and O.R.Zaine., ``Privacy Preservation when sharing data for clustering.,'' *In Proceedings of the International Workshop on Secure Data Management in a Connected World.*, 2004.

[10] Jaideep Vaidya and Clifton C.,``Privacy Preserving k-means clustering over vertically partitioned data.,'' *In Proceedings of ACM SIGKDD Conference*, pp.206-215,2003.

[11] Mohammad Ali Kadampur., Somayajulu D V L N., S.S.Shivaji Dhiraj.,G.P.Satyam.,``Privacy Preserving Clustering by Cluster Bulging for Information Sustenance.,'' *Proceedings of 5th International Conference on Information Automation for Sustenance (ICIAFS08)*, 2008.

[12] Vassillios S.Verykios.,Ahmed K. Elmagarmid., Elisa Bertino.,` `Association Rule Hiding.,'' *IEEE Transactions on Knowledge and Data Engineering.*,Vol.16, No.4,April 2004.

[13] S.J.Rizvi.,J.R.Haritsa., ``Privacy preserving association rule mining.,''

[14] *In the proceedings of 28th International Conference on Very Large Data Bases(VLDB)*., August 20-23 2002.

[15] Y.Lindell.,B.Pinkas.,``Privacy Preserving Data Mining.,'' *Journal of Cryptography.*,15(3):177-206,2002.

[16] Shuting Xu and Shuhua Lai ,``Fast Fourier transform based data perturbation method for privacy protection.,'' *In the proceedings of IEEE Conference on Intelligence and Security Informatics*, New Brunswick New Jersey, May 2007.

[17] Lian Liu.,Jie Wang.,Jun Zhang., ``Wavelet based data perturbation for simultaneous privacy preserving and statistics preserving.,'' *In Proceedings of IEEE International Conference on Data Mining workshop*., 2008.

[18] Charu C Agrwal.,Philip S Yu., " Privacy preserving data mining models and Algorithms.", *Springer Science+Business media.,LLC*..2008,

[19] J.R.Quinlan., " C4.5:Programs for Machine Learning.," Morgan Kaufmann,1993.

[20] Xiao Bai Li,Sumit Sarkar. ``Tree based data perturbation approach for Privacy Preserving Data Mining.,'' *In IEEE Transactions on Knowledge and Data Engineering,* Vol.18,No.9, September 2006.

[21] M.Bohanec and V.Rajkovic. "Knowledge acquisition and explanation for multi-attribute decision making" In 8th workshop on Expert Systems and their Applications,Avignon,France.pages 59-78,1988.

[22] Pinkas B."Cryptographic Techniques for Privacy-Preserving Data Mining" ACM SIGKDD Explorations, 4(2),2002

[23] Agrawal D. Aggarwal C.C. " On the Design and Quantification of Privacy Preserving Data mining algorithms." ACM PODS Conference, 2002.

[24] Fienberg S.E. and McIntyre J. "Data Swapping:Variations on a theme by Dalenius and Reiss." In Journal of Official Statistics, 21:309-323,2005.

[25] Muralidhar K. and Sarathy R. " Data Shuffling- a new masking approach for numerical data." Management Science, forthcoming, 2006.

[26] Y.Li,S.Zhu,L.Wang, and S.Jajodia " A privacy-enhanced microaggregation method" In Poc. Of 2nd International Symposium on Foundations of Information and Knowledge Systems, pp148-159,2002

[27] V.S. Iyengar."Transforming data to satisfy privacy constraints" In Proc. of SIGKDD'02, Edmonton, Alberta,Canada,2002.

[28] A.A.Hintoglu and Y.Saygin. " Suppressing microdata to prevent probabilistic classification based inference" In Proc. of Secure Data Management, 2nd VLDB workshop,SDM 2005 pp155-169, Trondheim Norway 2005.

[29] S.Rizvi, J.R. Harista " Maintaining data privacy in association rule mining" In Proc. of 28th VLDB Conference, pp682-693, Honk Kong China, 2002.

[30] Shuting Xu.,Shuhua Lai, "Fast Fourier Transform based data perturbation method for privacy protection" In Proc. of IEEE conference on Intelligence and Security Informatics, New Brunswick New Jersey, May 2007.

[31] Shibanth Mukharjee.,Zhiyuan Chen.,Arya Gangopadhyay "A privacy preserving technique for Euclidean distance-based mining algorithms using Fourier-related transforms" The VLDB journal 2006

[32] Lindell Y., Pinkas B."Privacy preserving Data Mining" CRYPTO 2000.

[33] Yu H.,Jiang X., Vaidya J."Privacy Preserving SVM using nonlinear Kernels on Horizontally partitioned Data. SAC Conference, 2006.

[34] Yu.H.,Vaidya J.,Jiang X."Privacy preserving SVM Classification on vertically partitioned data" PAKDD conference, 2006.

[35] Blum A.,Dwok C.,McSherry F., Nissim K. " Practical Privacy The SuLQ Framework" ACM PODS Conference, 2005

[36] Kenthapadi K.,Mishra N.,Nissim K.,"Simulatable Auditing" ACM PODS Conference 2005.

[37] Nabar S. Marthi B.,Kenthapadi K.,Mishra N., Motwani R.,"Towards Robustness in Querry Auditing" VLDB Conference, 2006.

[38] http://www.ies.uci.edu/~mlearn/MLRepository.html.archive

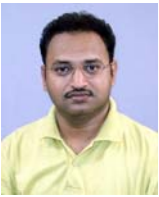

**Mohammad Ali Kadampur** is a research scholar (PhD) in the department of Computer Science and Engineering, National Institute of Technology Warangal A.P.India. He obtained his Bachelor of Engineering degree in Electronics and Communication from Karnataka University Dharwad in 1990, and worked in Industry and academia. He then obtained his Master of Technology in Computer Science and Engineering from Karnataka Regional Engineering College Surathkal India in 2001. He was awarded National Merit Scholarship by Govt of Karnataka . He has extensive teaching experience both in India and abroad for over 15 years and has published in leading International Conferences and journals. His current field of interests are privacy preserving data mining, Feature extraction and knowledge discovery. He is a member of Indian Association for Research in Computer Science (IARCS).

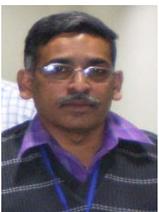

**Somayajulu D.V.L.N** is a professor in the Department of Computer Science and Engineering National Institute of Technology Warangal A.P India. He obtained his Master of Technology in Computer Science  from IIT Kharagpur in 1987 and PhD from IIT Delhi in 2002. He has been teaching in NIT Warangal since 1988 and has served from various academic positions. He has completed various consulting projects for Govt of Andhra Pradesh and Govt of India in his capacity as faculty of NIT Warangal. He was awarded the Engineer of the year in 2007 by Govt of Andhra Pradesh. He visited China and USA on the invitation of Oracle Corporation and Microsoft Corporation. He is a fellow of Institution of Engineers and member of IEEE. His current fields of interest are data warehousing, data mining and incomplete databases. He is a Fellow of IETE.